\documentclass[10pt,twocolumn]{article}

\usepackage[letterpaper,textwidth=7in,textheight=9.1in]{geometry}
\usepackage[T1]{fontenc}
\usepackage{newtxtext,newtxmath}
\usepackage{microtype}
\usepackage{amsmath}
\usepackage{booktabs}
\usepackage{balance}
\usepackage[font=small,labelfont=bf]{caption}
\usepackage{tikz}
\usepackage{xurl}
\usepackage{hyperref}
\usetikzlibrary{arrows.meta,fit,positioning}
\hypersetup{hidelinks}
\newcommand{\system}{LogDeck}
\newcommand{\alloy}{Grafana Alloy}
\newcommand{\fbit}{Fluent Bit}

\title{Source-Bounded Exact Recovery over Docker's Logs API}
\author{Kelvin Amoaba\\
{\small\href{mailto:kel.amoaba@gmail.com}{\texttt{kel.amoaba@gmail.com}}}}
\date{August 2026}

\begin{document}
\maketitle

\begin{abstract}
Reliable container-log acquisition is difficult because Docker can retain
records that a collector misses before attachment or during downtime. A
persisted read position does not by itself ensure recovery after lifecycle
changes. We study what exact recovery contract is achievable through Docker's
supported Logs API. We define
\emph{source-bounded exactness}: every retained, distinguishable source record
eventually appears exactly once in durable collector output. Our method uses a
generation-aware multiset oracle that separates source truncation from
collector omission and exposes simultaneous loss and replay. Applied to
\system{}, it uncovered a start-to-attachment race; a one-record attachment
overlap, finite Docker-API reconciliation, and exact insertion closed the
tested boundary.

We compare the fixed revision with unmodified \alloy{} 1.18.0, which uses the
same API and persists read positions; across 120 collector-runs, \system{} was
exact in 60/60 and Alloy in 20/60. Alloy succeeded at guarded startup and
process pause but omitted retained history when recovery required discovering
an exited or restarted source. In a causal control, a 5,000-record source
exited before collection: stock discovery was exact in 0/20 trials and acquired
nothing, while the same reader given the container ID recovered all records
exactly in 20/20. This reproduced on OrbStack and independent Ubuntu hosts with
Docker 29.4.0 and 24.0.9; both collectors recovered daemon restart, while
neither recovered records after source removal. Exactness assumes distinct
tuples of physical generation, timestamp, stream, and bytes; 200,000
byte-identical records across two drivers produced no observed collisions. Our
results show
that lifecycle reacquisition, not a persisted position alone, determines exact
recovery within the retained-source horizon. This is a bounded interface claim,
not a universal collector ranking or proof of collision freedom.
\end{abstract}

\section{Introduction}

An application writing a line to standard output does not imply that an
external collector durably acquired it. Between the write and a queryable
store are a logging driver, finite retained history, Docker's Logs API, a
followed HTTP response, collector buffers, and durable storage. These
boundaries do not fail in the same way.

Durable logs, checkpoints, and replay are established ideas. Chukwa places
recovery state near local sources and retransmits retained data~
\cite{rabkin2010chukwa}; Kafka provides a durable append-only log with
consumer-managed positions~\cite{kreps2011kafka}; and MillWheel combines
persistent state with fault-tolerant stream processing~
\cite{akidau2013millwheel}. The open engineering question here is narrower:
what recovery contract can a collector provide when its source is Docker's
supported Logs API rather than a durable broker or direct access to the
daemon's files?

The answer depends on more than whether a collector stores an offset. A
collector must first discover the physical container, attach to a stream,
recognize lifecycle changes, and enumerate retained history after a fault.
Those actions interact with a source whose contents and lifetime are controlled
by Docker. A stopped container can still have readable history; a removed
container does not. Our thesis is therefore precise: persisted read position
and lifecycle reacquisition are separate recovery obligations.

Measuring this boundary is also subtle. Suppose a collector misses one identity
and repeats another. Its output count is unchanged. Similarly, reusing sequence
numbers after a same-ID restart can make two physical generations appear to be
one. Existing collector benchmarks already study throughput, resource use,
rotation, and missing records~\cite{victoriametrics2026benchmark}. Our focus is
therefore not a generic collector ranking, but a duplicate-aware oracle and a
lifecycle fault model that can localize why acquisition is incomplete.

This study asks three questions:

\begin{enumerate}
  \item[\textbf{RQ1}] Can a Docker API follower converge to exactly one durable
        copy of every retained, distinguishable source record?
  \item[\textbf{RQ2}] Which attachment, collector, daemon, and container
        lifecycle boundaries cause omission or replay?
  \item[\textbf{RQ3}] Which mechanism is necessary for recovery, and where does
        the source lifetime make recovery impossible?
\end{enumerate}

The central result is not that one collector wins every condition. The Alloy
control succeeds at guarded startup, process pause, and live-restore daemon
restart. It fails when the tested schedule requires rediscovering an exited or
restarted source. A direct-target control then shows that the same Alloy reader
can read all retained history, localizing rapid-exit omission to target
construction rather than Docker's read path. A private-file \fbit{} control
also recovers exactly across collector restart and crash. These contrasts place
the contribution at the combination of supported interface, lifecycle
reconciliation, and exact oracle.

This paper makes four contributions:

\begin{enumerate}
  \item a source-bounded acquisition contract and a generation-aware multiset
        oracle that separately reports source truncation, omission, replay,
        unexpected identities, and malformed output;
  \item a Docker lifecycle suite spanning attachment, ephemeral exit,
        collector restart/crash/pause, same-ID restart, daemon restart, source
        rotation, and container removal;
  \item a concrete account of a start-to-attachment race found by the harness,
        together with a finite API-reconciliation design and controlled
        ablations of reconciliation and exact insertion; and
  \item comparative evidence from a supported-API baseline, a causal
        target-construction control replicated on two Linux environments, a
        private-file baseline, and two Docker logging drivers that identifies
        both recoverable and unrecoverable boundaries.
\end{enumerate}

\section{Model and Oracle}

\subsection{Log path and source horizon}

We model the acquisition path as

\[
\begin{aligned}
\text{stdout/stderr} &\rightarrow \text{driver history}
  \rightarrow \text{Docker Logs API}\\
&\rightarrow \text{collector buffer}
  \rightarrow \text{durable store}.
\end{aligned}
\]

The selected logging driver and its retention configuration determine whether
historical logs remain readable~\cite{docker2026logging}. No downstream
algorithm can recover a record after the source deletes the only retained
copy. The contract must therefore name the observation horizon explicitly.

\subsection{Source-bounded exactness}

Every generated line contains a run token, a logical generation, and a
sequence number. Let $G$ be the complete set of generated benchmark
identities. After the workload, the harness materializes Docker's retained
history; let $S \subseteq G$ be the set visible at that observation point.
Let $C$ be the collector output multiset, and let $m_C(r)$ be the observed
multiplicity of identity $r$.

The primary contract is

\[
\forall r \in S:\quad m_C(r)=1.
\]

The harness reports four disjoint questions rather than a single count:

\begin{itemize}
  \item $G\setminus S$: generated but absent from retained source;
  \item $\{r\in S:m_C(r)=0\}$: source-retained but collector-missing;
  \item $\sum_{r\in G}\max(0,m_C(r)-1)$: excess multiplicity; and
  \item $C\setminus G$: unexpected identities.
\end{itemize}

Malformed benchmark lines are reported separately. The same-ID scenario
increments the logical generation even though Docker reuses the physical
container ID. The oracle materializes all identities; it never infers loss
from a maximum sequence number or a total.

\subsection{Observable identity assumption}

The implementation does not receive the benchmark identity as privileged
metadata. It deduplicates the Docker-visible tuple

\[
(\text{physical generation},\ \text{timestamp},\ \text{stream},\
 \text{raw bytes}).
\]

Two legitimate records with the same four fields are observationally
indistinguishable at this interface. In the main workload, raw bytes contain
the benchmark generation and sequence, so every expected identity is
distinguishable. A separate stress test removes that convenience by emitting
byte-identical records and measuring whether Docker supplies distinct
timestamps. This tests an important ordinary case but cannot prove that a
collision is impossible on every driver, engine, or platform.

\section{Recovery Design and the Attachment Race}

\subsection{Finite reconciliation}

\system{} keeps the live path non-blocking. Followers feed bounded in-memory
queues, and a slow consumer cannot stall acquisition indefinitely. Queue
overwrite and lifecycle transitions produce reconciliation hints. Per-stream
watermarks record durable progress, while a startup snapshot prevents a newly
advanced live watermark from causing older retained history to be skipped.

A reconciliation pass rereads a finite Docker history interval with overlap
before the durable watermark or earliest gap. Inserts are idempotent under the
observable identity key. Gap hints are versioned: a pass may clear only the
version it observed, so a newer gap cannot be erased by stale recovery work.
Hints coalesce into an immediate wake, and periodic inventory remains a safety
net.

This is conventional at-least-once reread plus deduplication. The
Docker-specific issue is deciding \emph{which physical containers} and
\emph{which lifecycle intervals} must be reread through an API whose source
can outlive a running process but not container removal.

\subsection{A race exposed by the oracle}

The guarded runner found a schedule absent from the original randomized
matrix. At revision \texttt{b06a1ed}, an initial finite backfill completed
before the workload emitted its prefix. The \texttt{Tail: 0} live response
attached after that prefix. Because the producer then waited at a gate, no
later record arrived to trigger another recovery pass. The database contained
the discovered target and completed-backfill state but no target rows or
watermark.

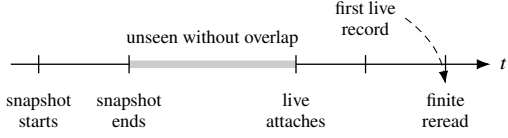
\begin{figure}[t]
\centering
\begin{tikzpicture}[
  x=0.92cm,y=1cm,
  >=Latex,
  every node/.style={font=\scriptsize,align=center}
]
  \draw[->,line width=0.5pt] (0.15,0.68) -- (7.05,0.68) node[right] {$t$};
  \draw[line width=2.8pt,gray!40] (1.85,0.68) -- (4.25,0.68);
  \node at (3.05,1.02) {unseen without overlap};
  \foreach \x in {0.55,1.85,4.25,5.25,6.40}
    \draw[line width=0.5pt] (\x,0.56) -- (\x,0.80);
  \node[below=4pt] at (0.55,0.56) {snapshot\\starts};
  \node[below=4pt] at (1.85,0.56) {snapshot\\ends};
  \node[below=4pt] at (4.25,0.56) {live\\attaches};
  \node[above=4pt] (first) at (5.25,0.80) {first live\\record};
  \node[below=4pt] (repair) at (6.40,0.56) {finite\\reread};
  \draw[->,densely dashed] (first.east) to[bend left=22] (repair.north);
\end{tikzpicture}
\caption{The start-to-attachment race and repair. A first-record attachment
overlap triggers a finite reread; exact insertion removes the repeated
boundary record.}
\label{fig:attachment}
\end{figure}

Focused failing tests reproduced this ordering. The repair asks each persistent
subscription for one retained record when opening the followed stream. The
first delivered record proves attachment and triggers one finite zero-point
reread. The reread overlaps the live boundary, and exact insertion absorbs the
intentional repeated record. Retried attachments apply the rule once per
attachment. Figure~\ref{fig:attachment} illustrates the schedule and repair.

\subsection{Bounded design claim}

Recovery is intended to converge when (1) the relevant record remains readable
from finite Docker history, (2) the collector can identify its physical
generation, (3) reconciliation eventually runs to completion, and (4)
distinct records have distinct observable identities. The evaluation tests
these assumptions and necessary mechanisms; it does not claim an unrestricted
exactly-once channel.

\section{Methodology}

\subsection{Black-box harness}

The standalone Go harness imports no \system{} package. It controls collectors
and producers through Docker and public HTTP endpoints. Every collector,
scenario, and trial receives a unique run token and target label. The expected
identity set is created before execution. After the workload, the harness
reads Docker history, snapshots collector output until exactness remains stable
for one second or an incomplete result plateaus, and writes a schema-versioned
JSON report. Raw sources and collector artifacts can be retained independently.

Fault boundaries are selected reproducibly between 20\% and 80\% of the
5,000-record workload. The seed depends on scenario and trial but not collector,
so both collectors receive the same ten boundaries. A \emph{collector-run} is
one collector executing one scenario once. All main-table denominators use
this unit.

\subsection{Collectors}

\paragraph{\system{}.}
The evaluated image is built from revision \texttt{58b3e4b} and pinned by
digest prefix \texttt{bd6b4630}. It includes the attachment repair. Research
ablation modes disable only reconciliation or exact insertion in the same
instrumented revision.

\paragraph{\alloy{} supported-API control.}
We run unmodified \alloy{} 1.18.0, digest prefix \texttt{491b0578}. Its
\texttt{discovery.docker} component supplies labeled targets to
\texttt{loki.source.docker}, which reads the Docker Logs API and stores read
offsets in a persistent positions file~\cite{grafana2026alloydocker}. A
\texttt{loki.echo} receiver makes records observable to the harness. The
configuration uses a 100\,ms discovery and source refresh interval. Before
guarded faults, the harness queries Alloy's component debug endpoint and
releases the producer only after the expected target reports a running reader.
Thus startup and pre-fault failures cannot be attributed to releasing the
workload before attachment.

\paragraph{Alloy target-construction control.}
To separate discovery from reading, a source emits all 5,000 records and exits
before either configuration starts. The stock configuration uses
\texttt{discovery.docker}; the control gives the exited container ID directly
to the same unmodified \texttt{loki.source.docker} component and image. The
harness stops the direct-target control after its first complete historical
transfer. This one-shot probe asks whether retained history is readable; it is
not proposed as a deployable discovery configuration, and it avoids measuring
subsequent retries of a permanently listed exited target.

Alloy's documented discovery arguments expose filters that limit discovered
resources, but no option to include stopped containers~
\cite{grafana2026alloydiscovery}. Version 1.18.0 delegates to Prometheus
Docker discovery~\cite{grafana2026alloydockersource}; its pinned v0.312.0
container-list call supplies filters but does not set the API's \texttt{All}
option~\cite{prometheus2026dockersdsource}.
Thus the direct-target condition is the configuration sweep available without
modifying Alloy: it bypasses target construction while holding its Docker
reader constant.

\paragraph{\fbit{} private-file control.}
Pinned \fbit{} 5.0.9 tails Docker's \texttt{json-file} records with a
persistent offset database. A Linux-process variant survives a real process
stop or \texttt{SIGKILL}. This is a useful stateful comparison, but it reads
the daemon's private log storage. Docker explicitly warns that these files are
designed for exclusive daemon access~\cite{docker2026jsonfile}; the privilege
and coupling difference is therefore part of the result.

\subsection{Faults}

Table~\ref{tab:faults} defines the principal schedules. Guarded scenarios
confirm attachment before their measured prefix. The rapid-exit scenario is
intentionally unguarded because discovery of an ephemeral producer is the
fault under test.

\begin{table}[t]
\caption{Principal fault schedules. Prefix and suffix refer to the seeded
fault boundary.}
\label{tab:faults}
\centering
\small
\begin{tabular}{@{}lp{5.2cm}@{}}
\toprule
Fault & Schedule \\
\midrule
Startup & Emit prefix; start collector; confirm attachment; emit suffix. \\
Rapid exit & Start collector; producer emits all records and exits immediately. \\
Exited source & Producer exits with all records retained; then start stock or direct-target reader. \\
Restart & Acquire prefix; stop collector; emit suffix and exit; restart collector. \\
Crash & Acquire prefix; \texttt{SIGKILL} collector; emit suffix and exit; restart. \\
Pause & Acquire prefix; pause process; emit suffix and exit; resume. \\
Same ID & Acquire generation 1; restart the same physical container for generation 2. \\
Daemon & Begin suffix while an isolated live-restore daemon is unavailable. \\
Removal & Verify full source, delete container, then restart collector. \\
\bottomrule
\end{tabular}
\end{table}

The daemon fault runs the harness inside a privileged, isolated
Docker-in-Docker engine so the ordinary Ubuntu daemon is untouched. A
supervisor restarts only the inner \texttt{dockerd}. Live restore is enabled
and verified before each matrix; Docker documents that this mode keeps
containers running while the daemon is unavailable~
\cite{docker2026liverestore}. The producer begins its suffix during the
outage. It may block if Docker's first-in, first-out (FIFO) logging buffer
fills, an explicit limitation of live restore.

For removal, the harness stops the collector after the prefix, releases and
waits for the complete producer, materializes and verifies all 5,000 source
identities, deletes the container, and restarts the collector. The report marks
this pre-removal oracle explicitly; the post-removal source horizon is empty.

\subsection{Environments}

The supported-API and removal matrices run inside the default Ubuntu OrbStack
machine on Linux/arm64 with eight CPUs and Docker Engine 29.1.3. The machine
has a separate Linux userspace and Docker daemon but shares the macOS host's
OrbStack kernel; it is not independent hardware. The daemon matrix uses the
same kernel with an isolated Docker 29.7.0 live-restore engine.

The target-construction control was run separately on (1) the OrbStack engine
with Docker 29.4.0 and kernel 7.0.11 and (2) an independent Ubuntu 22.04/arm64
cloud host with Linux 5.15 and Docker 24.0.9. The latter is a native daemon on
independent hardware. The version spread reflects experimental sequencing and
isolated-engine availability, not per-condition tuning: the main matrix ran
first on 29.1.3; focused target and same-ID controls ran later on the
then-current 29.4.0 OrbStack engine; the independent host remained pinned at
24.0.9; and the daemon runner used a separately pinned 29.7.0 image. Earlier
fixed-revision work and the \fbit{} process control also ran on that host. We
keep the environments separate rather than treating containerized Linux as a
second physical machine. The ablation matrix was rerun on August 1 in the
OrbStack environment at the pinned revision and seed reported in the artifact.

\section{Results}
\label{sec:results}

\subsection{The supported-API comparison localizes lifecycle recovery}

Table~\ref{tab:api} reports 120 collector-runs: two collectors, six scenarios,
and ten trials per cell. There were no operational errors. \system{} was
source-bounded exact in all 60 of its collector-runs. Alloy was exact in 20/60:
all startup and pause trials.

\begin{table}[t]
\caption{Supported Docker-API comparison, 5,000 records per collector-run.
Each cell is exact collector-runs out of ten.}
\label{tab:api}
\centering
\small
\begin{tabular}{@{}lrrl@{}}
\toprule
Fault & \system{} & Alloy & Alloy diagnostic \\
\midrule
Startup & 10/10 & 10/10 & known-target exact \\
Rapid exit & 10/10 & 0/10 & none acquired \\
Restart & 10/10 & 0/10 & pre-fault only \\
Crash & 10/10 & 0/10 & pre-fault only \\
Pause & 10/10 & 10/10 & reader catch-up \\
Same ID & 10/10 & 0/10 & generation 2 omitted \\
\midrule
Total & 60/60 & 20/60 & \\
\bottomrule
\end{tabular}
\end{table}

All ten Alloy rapid-exit runs acquired zero of 5,000 records even though the
post-run Docker oracle retained all 5,000. In restart and crash, Alloy's final
unique count equaled the randomized pre-fault boundary in every run; the
entire suffix was absent. The same was true after the same physical container
ID restarted for generation 2. Since Alloy succeeds when the existing reader
is paused, at guarded startup, and across daemon outage while the reader
survives, the main matrix separates continuing-reader success from lifecycle
reacquisition failure.

\subsection{Target construction, not the read path, causes rapid-exit omission}

Table~\ref{tab:target_control} holds Alloy's image, Docker reader, sink, and
source history constant while changing how the reader receives its target. In
every trial, Docker retained all 5,000 records before collection began. Stock
discovery acquired zero records in all 20 trials. Given the exited container
ID directly, the same unmodified reader acquired all 5,000 records exactly
once in all 20 trials. There were no operational errors.

\begin{table}[t]
\caption{Exited-source target-construction control. Each cell reports exact
collector-runs out of ten; parentheses report collector/source records per
run.}
\label{tab:target_control}
\centering
\small
\begin{tabular}{@{}llrr@{}}
\toprule
Environment & Docker & Stock discovery & Direct target \\
\midrule
OrbStack & 29.4.0 & 0/10 (0/5,000) & 10/10 (5,000/5,000) \\
Ubuntu & 24.0.9 & 0/10 (0/5,000) & 10/10 (5,000/5,000) \\
\bottomrule
\end{tabular}
\end{table}

The replication across a shared-kernel development engine and an independent
native Ubuntu engine closes the main environmental asymmetry. More
importantly, the control rejects the explanation that Alloy's Docker reader
cannot read exited-container history. The reader can; stock target discovery
does not present that exited source to it. This localizes the rapid-exit result
to target construction and supports the broader lifecycle-reacquisition thesis
without claiming that every Alloy deployment shares the same target graph.
Architecturally, the tested boundary arises because Alloy reuses Prometheus
Docker service discovery---whose native role is to feed scrape targets---to
drive historical log acquisition~\cite{prometheus2026configuration}. This
suggests a broader design check, not a universal failure claim: log pipelines
that inherit active-target discovery should verify whether retained history
remains reachable after targets exit.

\subsection{The oracle exposes a nondeterministic replay}

The multiset oracle found more than omission. Same-ID trial 6 retained 3,647
unique generation-1 identities, missed all 1,353 generation-2 identities, and
replayed \texttt{g1:1--1479}. Its final total was 5,126. A total-only method
would report 126 extra records and conceal both the missing suffix and 1,479
replays. We repeated the same ten seeded boundaries on Docker 29.4.0 while
preserving Alloy's final component state and positions. All ten reruns omitted
generation 2, including the same 3,647/1,353 split in trial 6, but none replayed
generation 1. The final component had no active reader and the persisted
positions map was empty in every rerun.

The replay is therefore nondeterministic under the otherwise identical fault
plan. We retain it because it demonstrates why multiplicity-aware measurement
is necessary, but we do not use it as a stable behavioral claim. The reproduced
same-ID finding is the complete omission of generation 2 after lifecycle
reuse.

\subsection{Daemon restart is a shared success}

The isolated live-restore matrix contains 20 collector-runs: ten per
collector, with the same ten randomized boundaries. Both \system{} and Alloy
were generated-set exact in 10/10, with no duplicate, unexpected, or malformed
identity and no operational error. The result rules out an overly broad claim
that Alloy cannot recover from Docker API disruption. When the collector
process and its in-memory state survive the daemon outage, the tested
configuration catches up exactly.

\subsection{Container removal marks the recovery horizon}

Immediately before each deletion, the independent oracle and Docker history
both contained all 5,000 generated identities. After deletion and collector
restart, both collectors ended with exactly the randomized pre-removal prefix
in all ten trials. Across 20 collector-runs, final unique count equaled the
fault boundary every time and duplicates were zero.

This is not a violation of source-bounded exactness: after removal, Docker no
longer exposes the suffix. It is direct evidence for the contract's limit. A
collector can preserve what it already acquired, but API reconciliation
cannot recover a source object that no longer exists.

\subsection{The harness found and closed a real race}

Before the attachment repair, the guarded runner repeatedly reached a state
with a completed initial backfill, an attached live stream, and no acquired
prefix. Focused tests forced that schedule. After the one-record overlap and
finite reread were added at revision \texttt{58b3e4b}, both the local and
independent fixed-revision matrices completed without the pre-fault timeout.
In the new API comparison, guarded startup was exact in 20/20 collector-runs,
including Alloy, confirming that the harness does not manufacture a startup
failure for the comparator.

This before/after sequence is a first-class result: the oracle and guarded
schedule exposed an acquisition hole that an unguarded burst benchmark had
not, and the repaired schedule remained exact across the evaluated
environments.

\subsection{Ablations identify two necessary mechanisms}

Table~\ref{tab:ablations} holds the revision and attachment overlap fixed while
disabling one mechanism. There were 80 collector-runs across the two paired
matrices. Every complete-design control was exact; every disabled condition
was inexact.

\begin{table}[t]
\caption{Fixed-revision ablations, ten collector-runs per cell. Ranges report
missing or excess multiplicity per run.}
\label{tab:ablations}
\centering
\small
\begin{tabular}{@{}lll@{}}
\toprule
Mechanism/fault & Complete & Disabled \\
\midrule
Reconcile/start & exact 10/10 & 1,458--3,498 missing \\
Reconcile/exit & exact 10/10 & 671--1,158 missing \\
Exact insert/start & exact 10/10 & 5,001--10,001 extra \\
Exact insert/restart & exact 10/10 & 1,805--6,464 extra \\
\bottomrule
\end{tabular}
\end{table}

Without reconciliation, startup missed exactly one fewer record than the
seeded prefix in every trial, consistent with the preserved one-record
attachment overlap. Without exact insertion, the collector acquired every
unique identity but retained repeated copies from overlapping rereads. These
results establish necessity under the tested faults, not sufficiency under
every Docker failure.

\subsection{A private-file baseline changes the interface tradeoff}

On the independent Linux process control, \system{} and \fbit{} were both
generated-set exact in all 40 collector-runs: ten graceful restarts and ten
\texttt{SIGKILL} trials per collector (Table~\ref{tab:filetail}). A broader
100-run supplemental matrix also found both exact across startup, rapid exit,
pause, same-ID restart, and source-horizon conditions.

\begin{table}[t]
\caption{Independent collector-process control, ten collector-runs per cell.}
\label{tab:filetail}
\centering
\small
\begin{tabular}{@{}lrr@{}}
\toprule
Fault & \system{} API & \fbit{} private file \\
\midrule
Graceful restart & 10/10 & 10/10 \\
\texttt{SIGKILL} & 10/10 & 10/10 \\
\bottomrule
\end{tabular}
\end{table}

The result prevents a claim that \system{} is uniquely exact. Stateful file
offsets are an effective recovery primitive here. The distinction is
deployment: \system{} uses Docker's supported API; this \fbit{} control
requires the isolated daemon's private \texttt{json-file} data root.

\subsection{Observable identity survives two tested drivers}

The identity stress emits the same raw stdout bytes 10,000 times per trial and
writes an independent bind-mounted oracle before each emission. The harness
reads Docker history with timestamps and counts collisions in the visible
tuple. Table~\ref{tab:identity} reports 20 trials and 200,000 records.

\begin{table}[t]
\caption{Byte-identical identity stress, ten collector-runs per driver.}
\label{tab:identity}
\centering
\small
\begin{tabular}{@{}lrrr@{}}
\toprule
Driver & Oracle/API & Tuple collisions & Exact \\
\midrule
\texttt{json-file} & 100,000/100,000 & 0 & 10/10 \\
\texttt{local} & 100,000/100,000 & 0 & 10/10 \\
\bottomrule
\end{tabular}
\end{table}

Every Docker-visible record had a distinct nanosecond timestamp, and
\system{} preserved exactly 100,000 records for each driver. This closes the
tested repeated-payload case on Docker 29.1.3. It does not close the deeper
case in which a driver or platform exposes two legitimate records with the
same timestamp, stream, bytes, and physical generation; such a collision would
silently collapse under the current key.

\subsection{Recovery work}

Direct instrumentation counted records returned by finite Docker rereads.
Across ten trials per condition, startup reread amplification was
1.29--1.70$\times$ (median 1.53$\times$), and collector restart was
1.23--1.65$\times$ (median 1.43$\times$). Exit burst required one full reread
in 9/10 trials and two in 1/10, for a median of 1.00$\times$ and maximum of
2.00$\times$. All 51 target recovery passes succeeded.

Amplification is target records returned by finite rereads divided by final
source-retained unique records. It is not database-write or byte
amplification. The 2.00$\times$ outlier demonstrates why exact insertion is
load-bearing even when recovery is otherwise complete. We omit the earlier
deployed CPU comparison because \system{} included an HTTP control plane,
parsing, and SQLite while the comparator used a file sink; that measurement
did not isolate the recovery mechanism.

\section{Related Work}

\paragraph{Reliable collection and durable transport.}
Chukwa is the closest conceptual foundation: agents checkpoint source state
and retransmit retained source-local records, while downstream duplicate
removal makes replay safe~\cite{rabkin2010chukwa}. Kafka provides durable
consumer replay~\cite{kreps2011kafka}, and MillWheel combines persistent state
with fault-tolerant stream processing~\cite{akidau2013millwheel}. These systems
preclude a novelty claim around replay or deduplication. Our contribution is
the Docker Logs API boundary, lifecycle suite, attachment race, and exact
oracle.

\paragraph{Container log pipelines.}
Prior systems assemble cluster logging from containerized open-source
components~\cite{singh2016cluster} or Docker pipelines around the
Elasticsearch, Logstash, and Kibana stack~\cite{liu2020docker}. Current
collector benchmarks include throughput, resources, rotation, discovery, and
missing-log tests~\cite{victoriametrics2026benchmark}. Our study complements
them with generation-aware multiplicity and controlled lifecycle boundaries.

\paragraph{File tailing and exporter resilience.}
\fbit{} persists file offsets and documents rotation behavior~
\cite{fluentbit2026tail}. OpenTelemetry Collector supports queues, retry, and a
persistent write-ahead log for exporter resilience~
\cite{opentelemetry2026resiliency}. These mechanisms address file replay or
downstream delivery; neither by itself guarantees that a Docker API source
discovered an ephemeral container. The Alloy result demonstrates the
difference within a supported API collector that already persists positions.

\paragraph{Storage and complete capture.}
Flight Data Recorder studies always-on capture of persistent-state
interactions~\cite{verbowski2006flight}, while Kellect targets efficient
lossless Windows kernel-event collection~\cite{chen2025kellect}. The compressed
log processor (CLP) addresses lossless compression and search after collection~
\cite{rodrigues2021clp}.
This paper studies the upstream question of whether intended records reached
durable collector storage at all.

\section{Scope and Threats to Validity}

\paragraph{Authorship and implementation bias.}
The author built \system{} and evaluates it. The study mitigates, but cannot
remove, this conflict with an external black-box harness, pinned images,
complete expected sets, preserved source observations, fixed-revision
ablations, and positive controls in which Alloy and \fbit{} match \system{}.
The strongest claim is a boundary result, not a universal collector ranking.

\paragraph{External validity.}
The full supported-API matrix covers one Docker 29.1.3 Linux/arm64 environment,
and the daemon matrix uses an isolated Docker 29.7.0 engine on the same kernel.
The causal target-construction control repeats on independent Ubuntu/arm64
hardware and Docker 24.0.9 as well as OrbStack Docker 29.4.0. This replication
supports localization of the rapid-exit result, not every cell of the larger
matrix. Results may differ across Docker versions, logging drivers, kernels,
platforms, collector configurations, and large fleets.

\paragraph{Fault and workload coverage.}
The suite covers attachment, rapid exit, collector restart/crash/pause,
same-ID restart, daemon restart with live restore, rotation, two-container
concurrency, stream interleaving, and container removal. It does not cover
host power loss, filesystem corruption, long outages that exhaust Docker's
live-restore FIFO, Kubernetes runtime interfaces, arbitrary multiline
parsers, or adversarially randomized schedules.

\paragraph{Identity.}
Exact insertion depends on physical generation, timestamp, stream, and bytes.
The two-driver repeated-payload experiment observed zero collisions in 200,000
records but is not a proof of timestamp injectivity. A collision is a silent
loss mode and remains an explicit contract exclusion.

\paragraph{Comparator configuration.}
Alloy 1.18.0 uses its documented Docker source, persistent data path, 100\,ms
refresh, label-filtered discovery, and a local echo receiver. The documented
discovery component has no option to enumerate stopped containers; its
version-pinned implementation lists containers without the Docker API's
\texttt{All} option. The direct-target control therefore bypasses discovery
rather than guessing at a nonexistent retention knob. Different target graphs
or custom discovery components may behave differently. The \fbit{} control
changes both source interface and privilege. The appendix's Vector control is
documented as best effort and is not evidence about Vector as a whole.

\section{Artifact and Reproducibility}

The ancillary archive contains the standalone Go harness, exact collector
configurations, the fixed-revision ablation patch and runner, JSON reports,
paper-facing summaries, environment and image provenance, verification
scripts, and cryptographic checksum manifests. The daemon runner creates a labeled,
privileged Docker-in-Docker engine, verifies live restore, imports the frozen
\system{} image, restarts only the inner daemon, and removes the isolated
engine afterward.

The principal reports contain 120 supported-API collector-runs, 40 exited-source
target-construction collector-runs across two environments, 10 same-ID
diagnostic reruns, 20 daemon collector-runs, 20 removal collector-runs, and 20
observable-identity collector-runs. The archive's verification script checks
denominators, fault-boundary equality, exactness, missing ranges, and duplicate
multiplicity. Large raw databases and logs are retained separately; compact
reports, manifests, harness code, and provenance accompany the paper.

\section{Conclusion}

Source-bounded exactness turns a vague losslessness claim into a falsifiable
contract: every record still readable from Docker must appear exactly once,
and records outside the source horizon are reported separately. The multiset
oracle found a real start-to-attachment race, verified its repair, and exposed
simultaneous omission and replay that totals conceal.

In the supported-API matrix, \system{} was exact in 60/60 collector-runs and
Alloy in 20/60. Alloy's successes under startup, pause, and daemon restart make
the negative result specific: the tested weakness appears when lifecycle
recovery requires reacquiring a source that has already exited or restarted.
The direct-target control provides the causal distinction: Alloy's unmodified
reader recovered all retained rapid-exit history in 20/20 trials, while stock
discovery presented none of it in 20/20 trials across two environments. A
private-file \fbit{} control also recovered exactly, at the cost of coupling to
Docker's internal storage. When the source container was removed, neither
collector recovered the suffix. Together these results identify the useful
claim: finite API reconciliation plus exact insertion can recover retained,
distinguishable Docker records across the tested lifecycle faults, but no
persisted position can replace reacquisition, outlive the source, or
distinguish information Docker does not expose.

\appendix

\section{Best-Effort API Control}

An earlier independent-cloud matrix compared the fixed \system{} revision with
pinned Vector 0.55.0 across six scenarios and ten trials per collector.
\system{} was exact in 60/60 collector-runs; the configured Vector Docker
source was inexact in 60/60, principally through final-record duplication.
Focused size curves and raw JavaScript Object Notation Lines (JSONL) localized
those duplicates to repeated byte-equivalent events. Vector documents this
source as best effort~\cite{vector2026dockerlogs}; consequently we treat it as a harness sensitivity
control rather than the paper's main comparator.

\section{Forward Transport Control}

A separate experiment changed the source interface to Docker's
\texttt{fluentd} driver and a native \fbit{} Forward input. When the producer
remained alive through collector restart, output matched all 50,000 oracle
identities in 10/10 collector-runs. When the producer exited before reconnect,
every one of 27 collector-runs retained only its pre-fault prefix, independent
of tested asynchronous buffer limits from 128 to 1,048,576 events.

This result concerns transport lifetime rather than Logs-API acquisition.
Docker's in-memory driver queue can bridge a collector restart for a continuing
producer, but not producer/logger destruction before drain
~\cite{docker2026fluentd,fluentbit2026forward}.

\balance
{\footnotesize
\bibliographystyle{plain}
\bibliography{references}
}

\end{document}